\documentclass[conference]{IEEEtran}
\usepackage{multicol}
\usepackage{lipsum,graphicx}
\usepackage{float}
\usepackage{epstopdf}
\usepackage{ifthen}
\usepackage[latin1]{inputenc}

\usepackage{amssymb}
\usepackage[cmex10]{amsmath}
\usepackage{dcolumn}
\usepackage[nolist]{acronym}

\DeclareGraphicsExtensions{.pdf,.jpeg,.png,.eps}
\usepackage{booktabs}

\usepackage[caption=false,font=footnotesize]{subfig}
\usepackage{comment}
\usepackage{color}
\usepackage[table]{xcolor}
\usepackage{cite}

\usepackage{setspace}
\usepackage{mathrsfs}          
\usepackage{amsmath, amssymb, amsthm}
\usepackage{amsfonts}
\usepackage{mathrsfs}
\usepackage{multirow}        
\usepackage{framed}
\usepackage{psfrag}
\usepackage{graphicx}
\usepackage{textcomp, gensymb} 
\usepackage{algpseudocode} 

\newcolumntype{d}[1]{D{.}{\cdot}{#1} }

\usepackage{bm}
\usepackage{mathtools}
\usepackage{siunitx}

\newcommand{\PreserveBackslash}[1]{\let\temp=\\#1\let\\=\temp}
\newcolumntype{C}[1]{>{\PreserveBackslash\centering}p{#1}}
\newcolumntype{R}[1]{>{\PreserveBackslash\raggedleft}p{#1}}
\newcolumntype{L}[1]{>{\PreserveBackslash\raggedright}p{#1}}

\newcommand{\figref}[1]{Fig.~\ref{#1}}

\usepackage{lipsum}
\usepackage{mathtools}
\usepackage{cuted}
\usepackage{stmaryrd}

\usepackage{tabularx}
\usepackage{tabulary}
\usepackage{tikz}
\usepackage{pgfplots}
\usepgfplotslibrary{units}
\pgfplotsset{compat=1.18}

\usepackage{algpseudocode}
\usepackage{algorithm}

\begin{document}
	\begin{acronym}
\acro{1G}{first generation}
\acro{2G}{second generation}
\acro{3G}{third generation}
\acro{3GPP}{Third Generation Partnership Project}
\acro{4G}{fourth generation}
\acro{5G}{fifth generation}
\acro{6G}{sixth-generation}
\acro{802.11}{IEEE 802.11 specifications}
\acro{A/D}{analog-to-digital}
\acro{ADC}{analog-to-digital}
\acro{AGC}{automatic gain control}
\acro{AM}{amplitude modulation}
\acro{AoA}{angle of arrival}
\acro{AP}{access point}
\acro{AI}{artificial intelligence}
\acro{AR}{augmented reality}
\acro{ASIC}{application-specific integrated circuit}
\acro{ASIP}{Application Specific Integrated Processors}
\acro{AWGN}{additive white Gaussian noise}
\acro{BB}{base-band}
\acro{BCJR}{Bahl, Cocke, Jelinek and Raviv}
\acro{BER}{bit error rate}
\acro{BFDM}{bi-orthogonal frequency division multiplexing}
\acro{BPSK}{binary phase shift keying}
\acro{BS}{base stations}
\acro{CA}{carrier aggregation}
\acro{CAF}{cyclic autocorrelation function}
\acro{Car-2-x}{car-to-car and car-to-infrastructure communication}
\acro{CAZAC}{constant amplitude zero autocorrelation waveform}
\acro{CB-FMT}{cyclic block filtered multitone}
\acro{CCDF}{complementary cumulative density function}
\acro{CDF}{cumulative density function}
\acro{CDMA}{code-division multiple access}
\acro{CFO}{carrier frequency offset}
\acro{CIR}{channel impulse response}
\acro{CM}{complex multiplication}
\acro{COFDM}{coded-\acs{OFDM}}
\acro{CoMP}{coordinated multi point}
\acro{COQAM}{cyclic OQAM}
\acro{CP}{cyclic prefix}
\acro{CRB}{Cramer-Rao bound}
\acro{CPE}{constant phase error}
\acro{CR}{cognitive radio}
\acro{CRC}{cyclic redundancy check}
\acro{CRLB}{Cram\'{e}r-Rao lower bound}
\acro{CS}{cyclic suffix}
\acro{CSI}{channel state information}
\acro{CSMA}{carrier-sense multiple access}
\acro{CWCU}{component-wise conditionally unbiased}
\acro{D/A}{digital-to-analog}
\acro{D2D}{device-to-device}
\acro{DAC}{digital-to-analog converter}
\acro{DBF}{digital beamforming}
\acro{DC}{direct current}
\acro{DFE}{decision feedback equalizer}
\acro{DFT}{discrete Fourier transform}
\acro{DL}{downlink}
\acro{DMT}{discrete multitone}
\acro{DNN}{deep neural network}
\acro{DSA}{dynamic spectrum access}
\acro{DSL}{digital subscriber line}
\acro{DSP}{digital signal processor}
\acro{DTFT}{discrete-time Fourier transform}
\acro{DVB}{digital video broadcasting}
\acro{DVB-T}{terrestrial digital video broadcasting}
\acro{DWMT}{discrete wavelet multi tone}
\acro{DZT}{discrete Zak transform}
\acro{E2E}{end-to-end}
\acro{eNodeB}{evolved node b base station}
\acro{E-SNR}{effective signal-to-noise ratio}
\acro{EVD}{eigenvalue decomposition}
\acro{EVM}{error vector magnitude}
\acro{FBMC}{filter bank multicarrier}
\acro{FD}{frequency-domain}
\acro{FDA}{frequency diverse array}
\acro{FDD}{frequency-division duplexing}
\acro{FDE}{frequency domain equalization}
\acro{FDM}{frequency division multiplex}
\acro{FDMA}{frequency-division multiple access}
\acro{FEC}{forward error correction}
\acro{FER}{frame error rate}
\acro{FFT}{fast Fourier transform}
\acro{FI}{Fisher information}
\acro{FIR}{finite impulse response}
\acro{FM}		{frequency modulation}
\acro{FMT}{filtered multi tone}
\acro{FO}{frequency offset}
\acro{F-OFDM}{filtered-\acs{OFDM}}
\acro{FPGA}{field programmable gate array}
\acro{FSC}{frequency selective channel}
\acro{FS-OQAM-GFDM}{frequency-shift OQAM-GFDM}
\acro{FT}{Fourier transform}
\acro{FTD}{fractional time delay}
\acro{FTN}{faster-than-Nyquist signaling}
\acro{GFDM}{generalized frequency division multiplexing}
\acro{GFDMA}{generalized frequency division multiple access}
\acro{GMC-CDM}{generalized	multicarrier code-division multiplexing}
\acro{GNSS}{global navigation satellite system}
\acro{GPSDO}{GPS disciplined oscillator}
\acro{GPS}{global positioning system}
\acro{GS}{guard symbols}
\acro{GSM}{Groupe Sp\'{e}cial Mobile}
\acro{GUI}{graphical user interface}
\acro{H2H}{human-to-human}
\acro{H2M}{human-to-machine}
\acro{HF}{high frequency}
\acro{HPBW}{half-power beam-width}
\acro{HTC}{human type communication}
\acro{I}{in-phase}
\acro{i.i.d.}{independent and identically distributed}
\acro{IB}{in-band}
\acro{IBI}{inter-block interference}
\acro{IC}{interference cancellation}
\acro{ICI}{inter-carrier interference}
\acro{ICT}{information and communication technologies}
\acro{ICV}{information coefficient vector}
\acro{IDFT}{inverse discrete Fourier transform}
\acro{IDMA}{interleave division multiple access}
\acro{IEEE}{institute of electrical and electronics engineers}
\acro{IF}{intermediate frequency}
\acro{IFFT}{inverse fast Fourier transform}
\acro{IoT}{Internet of Things}
\acro{IOTA}{isotropic orthogonal transform algorithm}
\acro{IP}{internet protocole}
\acro{IP-core}{intellectual property core}
\acro{ISDB-T}{terrestrial integrated services digital broadcasting}
\acro{ISDN}{integrated services digital network}
\acro{ISI}{inter-symbol interference}
\acro{ITU}{International Telecommunication Union}
\acro{IUI}{inter-user interference}
\acro{JCnS}{joint communication and sensing}
\acro{KDE}{kernel density estimate}
\acro{LAN}{local area netwrok}
\acro{LLR}{log-likelihood ratio}
\acro{LMMSE}{linear minimum mean square error}
\acro{LNA}{low noise amplifier}
\acro{LO}{local oscillator}
\acro{LOS}{line-of-sight}
\acro{LoS}{line of sight}
\acro{LP}{low-pass}
\acro{LPF}{low-pass filter}
\acro{LS}{least squares}
\acro{LTE}{long term evolution}
\acro{LTE-A}{LTE-Advanced}
\acro{LTIV}{linear time invariant}
\acro{LTV}{linear time variant}
\acro{LUT}{lookup table}
\acro{M2M}{machine-to-machine}
\acro{MA}{multiple access}
\acro{MAC}{multiple access control}
\acro{MAP}{maximum a posteriori}
\acro{MAE}{mean absolute error}
\acro{MC}{multicarrier}
\acro{MCA}{multicarrier access}
\acro{MCM}{multicarrier modulation}
\acro{MCS}{modulation coding scheme}
\acro{MF}{matched filter}
\acro{MF-SIC}{matched filter with successive interference cancellation}
\acro{MIMO}{multiple-input multiple-output}
\acro{MISO}{multiple-input single-output}
\acro{ML}{maximum likelihood}
\acro{MLD}{maximum likelihood detection}
\acro{MLE}{maximum likelihood estimator}
\acro{MMSE}{minimum mean squared error}
\acro{MRC}{maximum ratio combining}
\acro{MS}{mobile stations}
\acro{MSE}{mean squared error}
\acro{MSK}{Minimum-shift keying}
\acro{MSSS}[MSSS]	{mean-square signal separation}
\acro{MTC}{machine type communication}
\acro{MU}{multi user}
\acro{MVDR}{minimum variance distortionless response}
\acro{MVUE}{minimum variance unbiased estimator}
\acro{NEF}{noise enhancement factor}
\acro{NLOS}{non-line-of-sight}
\acro{NMSE}{normalized mean-squared error}
\acro{NOMA}{non-orthogonal multiple access}
\acro{NPR}{near-perfect reconstruction}
\acro{NRZ}{non-return-to-zero}
\acro{OU}{Ornstein-Uhlenbeck}
\acro{OFDM}{orthogonal frequency division multiplexing}
\acro{OFDMA}{orthogonal frequency division multiple access}
\acro{OMP}{orthogonal matching pursuit}
\acro{OOB}{out-of-band}
\acro{OQAM}{offset quadrature amplitude modulation}
\acro{OQPSK}{offset quadrature phase shift keying}
\acro{OTA}{Over-the-Air}
\acro{OTFS}{orthogonal time frequency space}
\acro{PA}{power amplifier}
\acro{PAM}{pulse amplitude modulation}
\acro{PAPR}{peak-to-average power ratio}
\acro{PB}{pass-band}
\acro{PC-CC}{parallel concatenated convolutional code}
\acro{PCP}{pseudo-circular pre/post-amble}
\acro{PD}{probability of detection}
\acro{pdf}{probability density function}
\acro{PDF}{probability distribution function}
\acro{PDP}{power delay profile}
\acro{PFA}{probability of false alarm}
\acro{PFD}{phase-frequency detector}
\acro{PHY}{physical layer}
\acro{PIC}{parallel interference cancellation}
\acro{PLC}{power line communication}
\acro{PLL}{phase-locked loop}
\acro{PMF}{probability mass function}
\acro{PN}{pseudo noise}
\acro{ppm}{parts per million}
\acro{PRB}{physical resource block}
\acro{PRB}{physical resource block}
\acro{PSD}{power spectral density}
\acro{Q}{quadrature-phase}
\acro{QAM}{quadrature amplitude modulation}
\acro{QoS}{quality of service}
\acro{QPSK}{quadrature phase shift keying}
\acro{R/W}{read-or-write}
\acro{RAM}{random-access memmory}
\acro{RAN}{radio access network}
\acro{RAT}{radio access technologies}
\acro{RC}{raised cosine}
\acro{RF}{radio frequency}
\acro{rms}{root mean square}
\acro{RMS}{root mean square}
\acro{RMSE}{root mean square error}
\acro{RSSI}{received signal strength indicator}
\acro{RRC}{root raised cosine}
\acro{REF}{reference}
\acro{RW}{read-and-write}
\acro{RX}{receiver}
\acro{SC}{single-carrier}
\acro{SCA}{single-carrier access}
\acro{SC-FDE}{single-carrier with frequency domain equalization}
\acro{SC-FDM}{single-carrier frequency division multiplexing}
\acro{SC-FDMA}{single-carrier frequency division multiple access}
\acro{SD}{sphere decoding}
\acro{SDD}{space-division duplexing}
\acro{SDMA}{space division multiple access}
\acro{SDR}{software-defined radio}
\acro{SDW}{software-defined waveform}
\acro{SEFDM}{spectrally efficient frequency division multiplexing}
\acro{SE-FDM}{spectrally efficient frequency division multiplexing}
\acro{SER}{symbol error rate}
\acro{SIC}{successive interference cancellation}
\acro{SINR}{signal-to-interference-plus-noise ratio}
\acro{SIR}{signal-to-interference ratio}
\acro{SISO}{single-input, single-output}
\acro{SMS}{Short Message Service}
\acro{SNR}{signal-to-noise ratio}
\acro{STC}{space-time coding}
\acro{STFT}{short-time Fourier transform}
\acro{STO}{sample-time-offset}
\acro{SU}{single user}
\acro{SVD}{singular value decomposition}
\acro{TX}{transmitter}
\acro{TD}{time-domain}	
\acro{TDD}{time-division duplexing}
\acro{TTD}{true-time delay}
\acro{TDMA}{time-division multiple access}
\acro{TFL}{time-frequency localization}
\acro{TO}{time offset}
\acro{ToA}{time of arrival}
\acro{TDoA}{time difference of arrival}
\acro{TS-OQAM-GFDM}{time-shifted OQAM-GFDM}
\acro{UE}{user equipment}
\acro{UFMC}{universally filtered multicarrier}
\acro{UL}{uplink}
\acro{ULA}{uniform linear array}
\acro{US}{uncorrelated scattering}
\acro{USB}{universal serial bus}
\acro{USRP}{universal software radio peripheral}
\acro{UW}{unique word}
\acro{VLC}{visible light communications}
\acro{VR}{virtual reality}
\acro{VCO}{voltage-controlled oscillator}
\acro{WCP}{windowing and \acs{CP}}	
\acro{WHT}{Walsh-Hadamard transform}
\acro{WiMAX}{worldwide interoperability for microwave access}
\acro{WLAN}{wireless local area network}
\acro{W-OFDM}{windowed-\acs{OFDM}}	
\acro{WOLA}{windowing and overlapping}	
\acro{WSS}{wide-sense stationary}
\acro{ZCT}{Zadoff-Chu transform}
\acro{ZF}{zero-forcing}
\acro{ZMCSCG}{zero-mean circularly-symmetric complex Gaussian}
\acro{ZP}{zero-padding}
\acro{ZT}{zero-tail}
\acro{URLLC}{ultra-reliable low-latency communications}
\acro{UDP}{user datagram protocol}
\acro{HSI}{human system interface}
\acro{HMI}{human machine interface}
\acro{VR} {visual reality} 
\acro{AGV}{automated guided vehicles}
\acro{MEC}{multiaccess edge cloud}
\acro{TI} {tactile Internet}
\acro{IMT}{ international mobile telecommunications}
\acro{GN}{gateway node}
\acro{CN}{control node}
\acro{NC}{network controller}
\acro{SN}{sensor node}
\acro{AN}{actuator node}
\acro{HN}{haptic node}
\acro{TD}{tactile devices}
\acro{SE}{supporting engine}
\acro{AI}{artificial intelligence}
\acro{TSM}{tactile service manager}
\acro{TTI}{transmission time interval}
\acro{NR}{new radio}
\acro{SDN}{software defined networking}
\acro{NFV}{ network function virtualization}
\acro{CPS}{cyber-physical system}
\acro{TSN}{Time-Sensitive Networking}
\acro{FEC}{forward error correction}
\acro{STC}{space-time  coding}
\acro{HARQ}{hybrid automatic repeat request}
\acro{CoMP} {Coordinated multipoint}
\acro{HIS}{human system interface }
\acro{RU}{radio unit}
\acro{CU}{central unit}
\acro{AoD} {angle of departure}
\end{acronym}
    
    
    \title{Experimental Comparison of Local and Over-the-Air Phase Calibration for MIMO Arrays}
    
	
	\author{
		\IEEEauthorblockN{
			Carl Collmann, Ahmad Nimr, Gerhard Fettweis 
			}
			
		\IEEEauthorblockA{
		Vodafone Chair Mobile Communications Systems, Technische Universit\"{a}t Dresden, Germany\\ \small\texttt{\{carl.collmann, ahmad.nimr,  gerhard.fettweis\}@tu-dresden.de}\\
		}
		}
	\maketitle
	\IEEEpeerreviewmaketitle
	
\begin{abstract}
Communication performance and channel estimation accuracy in \ac{MIMO} systems are known to be limited by hardware impairments.
Specifically, the presence of phase impairments, such as phase noise, makes real-time coherent transmission a challenging task.
While phase impairment compensation is typically performed at the receiver, practical methods for enabling coherent transmission at the transmitter side remain underexplored.
Established methods for \ac{OTA} calibration of \ac{MIMO} systems face several limitations such as assumptions of phase stationarity and accurate channel knowledge.
In this work, a real-time local phase calibration method is experimentally compared with \ac{OTA} calibration on a fully digital array of \ac{USRP} X310 software-defined radios.
Using \ac{RMS} cycle-to-cycle jitter as a metric, it is shown that for low and high synchronization signal bandwidths, both approaches effectively eliminate phase drift and whiten the phase noise.
Local calibration achieves higher phase stability and is channel-independent, whereas \ac{OTA} calibration requires no additional hardware but is sensitive to multipath effects and channel-induced impairments.
Practical deployment trade-offs are discussed based on the measurement results.
\end{abstract}

 
	\acresetall
\section{Introduction}\label{sec:introduction}

Hardware impairments in \ac{MIMO} mobile communication systems are known to limit channel capacity and channel estimation accuracy \cite{6891254}.
The presence of residual transmitter impairments reduces the achievable data rate in \ac{MIMO} systems and scales with transmitter power and number of antennas \cite{5782106}.
Specifically, the presence of phase noise reduces the maximum gain of maximum-ratio combining for non-coherent operation \cite{6902790}, reflecting the situation of a fully digital array.
Furthermore, for a high number of antennas, the presence and magnitude of phase noise necessitates periodic calibration \cite{6902790}, which is consistent with the findings of our previous works \cite{Coll202606}.
It has been shown that, under the condition of phase noise stationarity, compensation of the \ac{CPE} after channel estimation can mitigate the phase noise impact at the receiver \cite{9419749}.
This, however, necessitates short data transmission intervals to maintain the assumption of phase stationarity and does not enable coherent transmission.  

Established phase synchronization approaches for \ac{MIMO} transceivers \cite{rogalin2015,5361473,7839181,3gpp_nr_ptrs_chapter_7_4} typically rely on \ac{OTA} transmission and feedback between transmitter and receiver.
A known channel or used location is often assumed and these approaches employ a multiplexing scheme such as \ac{TDMA} to isolate the channels observed at each antenna.
The calibration scheme introduced in \cite{rogalin2015} uses uplink pilot signals to pre-distort downlink signals, enabling coherent combining.
The approach faces the limitation of not separating the channel from \ac{RF} chain phase and relies on channel reciprocity, which may be violated when significant phase drift occurs over longer transmission intervals.
The method for coherent transceiver design \cite{7839181} follows a similar approach to \cite{rogalin2015}, where channel coefficients are estimated and applied as precoding for equalization.
Other works on transmit beamforming utilize feedback from the receiver to calibrate the transmitter \cite{5361473}, similar to \cite{Coll202606} where a local reference chain replaces the \ac{OTA} feedback link.
While the synchronization method described in \cite{5361473} converges, it is limited by the assumption of stationary phase disturbances and requiring many iterations to achieve sufficient coherence.
To obtain the precise synchronization required for \ac{MIMO} radar imaging, works like \cite{9136646} assume a fixed known geometrical reference point for calibration.
Most established \ac{OTA} phase synchronization methods are sensitive to propagation channel effects, such as multipath delay spread, which can degrade phase estimation accuracy.
Moreover, many existing phase synchronization techniques have only been evaluated through simulation, with limited experimental validation.

To overcome these limitations, we previously introduced \cite{Coll202606} a simple method for phase calibration of a transmitting fully digital \ac{ULA}, using a local \ac{RF} chain as reference and provided validation with measurements.
In this work, we extend the previous local calibration study to an \ac{OTA} calibration setup, thereby enabling a direct comparison of the two approaches and linking our method to conventional synchronization techniques \cite{rogalin2015,5361473,3gpp_nr_ptrs_chapter_7_4}.
The performance of local and \ac{OTA} synchronization is compared in terms of their \ac{RMS} cycle-to-cycle jitter and validated with measurements \cite{1yjg-9863-26}.
In \cite{Coll202606}, the calibration method was verified only for a low bandwidth synchronization signal.
A further key objective of the present work is therefore to compare calibration performance for both low and high bandwidth signals and to demonstrate effectiveness in both scenarios.

The remainder of this paper is organized as follows.
Section \ref{sec:system_model} introduces the system model for local and \ac{OTA} phase estimation.
Section \ref{sec:impairment_model} describes the measurement setup.
Measurement results are shown in section \ref{sec:meas_result} to validate the calibration approach.
Section \ref{sec:evaluation} discusses the advantages of local vs \ac{OTA} calibration to provide practical insights.
The paper concludes with key results in section \ref{sec:conclusions}.

\section{System Model}\label{sec:system_model}

\begin{figure}[tb]
    	\centering
    	\includegraphics[width=\linewidth]{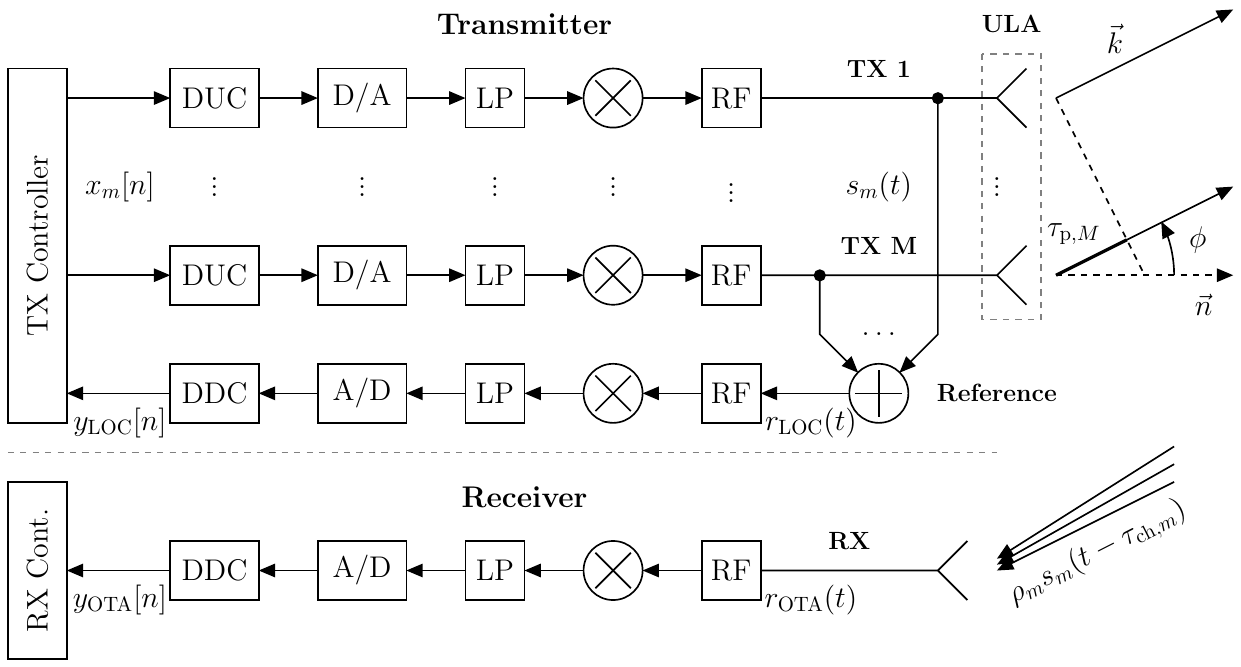}
        \caption{System model with local and \ac{OTA} receiver for phase calibration of transmitting \ac{ULA}.}
    	\label{fig:TX_sync_setup}
\end{figure}

\subsection{Transmit Signal Model}

A mobile communication system is considered with a \ac{ULA} of $M$ elements, as illustrated in \figref{fig:TX_sync_setup}.
The key objective is to calibrate the phases of the transmitted signals to allow for coherent transmission.
This can be achieved by observing the phases of the respective transmitting chains at a common reference point; either a co-located receiver, or an \ac{OTA} receiver. 
The phase estimates for the TX \ac{RF} chains obtained at the receivers are fed back to the TX controller for precoding to achieve coherent transmission.
It is assumed that a common frequency reference signal ($\SI{10}{MHz}$) and trigger signal (e.g., 1-pulse-per-second) are available (\ac{GPSDO} disciplined oscillators at transmitter and receiver).
The \ac{RF} chain and frontend components are assumed to have variable phase responses.
The observed phase offsets are assumed to be caused by variable phase responses of \ac{RF} components, phase noise effects of the oscillators, \ac{CFO} for the synthesized carriers and, in the \ac{OTA} case, additive white Gaussian noise and propagation delays that affect the channel phase.

Each TX chain $m$ transmits a chirp $x[n]$ of $N$ samples duration as a synchronization signal in a \ac{TDMA} scheme. 
The baseband signal after D/A conversion and low-pass filtering is $x_\text{BB}(t)=e^{j\pi \frac{B}{T}t^2}$, where $B$ refers to the bandwidth of the signal and $T=N T_\text{s}$ to its duration with sample duration $T_\text{s}$.
It is assumed that the power of $x_\text{BB}(t)$ is normalized to $1$.
After up-conversion, the signal in bandpass domain for chain $m$ is
\begin{align}
    x_{\text{BP},m}(t) = x_\text{BB}(t) e^{j(2\pi f_{\text{c},m}t + \theta_{\text{OS},m}(t))},
\end{align}
with $f_{\text{c},m}$ the synthesized carrier frequency and $\theta_{\text{OS},m}(t)$ the time-dependent oscillator phase.
Including the constant frontend phase shift $\theta_{\text{RF},m}$, the transmitted \ac{TDMA} signal becomes
\begin{align}
    s_{m}(t) = x_{\text{BB}}(t - mT) e^{j (2\pi f_{\text{c},m}(t - mT) + \theta_{\text{OS},m}(t-mT) +\theta_{\text{RF},m})}.
    \label{eq:TX_sync_PB}
\end{align}
This shift is caused by the phase response of amplifiers, switches, splitters, filters, and other \ac{RF} frontend components.

\subsection{Receive Signal Model}

The received passband signal at the local reference \ac{RF} chain is the sum of the transmit signals
\begin{align}
    r_\text{LOC}(t) = \sum_{m=1}^{M} s_{m}(t).
\end{align}
After downconversion and rectangular windowing to isolate the $m$-th \ac{TDMA} slot, the baseband signal is
\begin{align}
    y_{\text{LOC},m}(t) = s_{m}(t) e^{-j(2\pi f_{\text{c,LOC}}t + \phi_{\text{OS}}(t) + \phi_{\text{RF}})},
\end{align}
where $\phi_{\text{OS}}(t)$ refers to the phase of the time-dependent oscillator signal, $\phi_{\text{RF}}$ to the phase shift caused by the \ac{RF} frontend and $f_{\text{c,LOC}}$ to the synthesized carrier frequency at the local reference receiver chain.

At the \ac{OTA} receiver, the passband signal is affected by the channel:
\begin{align}
    r_\text{OTA}(t) = \sum_{m=1}^{M} \rho_m s_{m}(t-\tau_{\text{p},m}-\tau_{\text{ch},m}) + v(t).
\end{align}
Parameter $\rho_m$ refers to the complex channel gain, $\tau_{\text{p},m}=m\frac{d}{c}\sin \phi$ to the propagation delay, for \ac{ULA} element spacing $d$, speed of light $c$ and direction of RX relative to \ac{ULA} array normal $\phi$, path delay $\tau_{\text{ch},m}$ and Gaussian noise $v(t) \sim \mathcal{N}(0, \sigma_v^2)$.
For simplicity it is assumed that $\rho_m\approx\rho$ and equal path delays $\tau_{\text{ch},m}\approx \tau_{\text{ch}}$, which is appropriate for \ac{LOS} and far-field conditions.
The windowed baseband signal for slot $m$ is then
\begin{align}
    &y_{\text{OTA},m}(t) = \\
    &\rho s_{m}(t-\tau_{\text{p},m}-\tau_{\text{ch}}) e^{-j(2\pi f_\text{c,OTA}t + \varphi_{\text{OS}}(t) + \varphi_{\text{RF}})} + v'(t),\nonumber
\end{align}
with $\varphi_{\text{OS}}(t)$ referring to the phase of the time-dependent oscillator signal, $\varphi_{\text{RF}}$ to the phase shift caused by the \ac{RF} frontend, $f_\text{c,OTA}$ to the synthesized carrier frequency and $v'$ representing the baseband equivalent noise observed at the \ac{OTA} receiver chain.

The system function for \ac{RF} chain $m$ is obtained by multiplication of the received baseband signals with the complex conjugate of the known transmitted baseband signal.
The system functions for the local \ac{RF} chain and the \ac{OTA} receiver are
\begin{align}
    &\hat{h}_{\text{LOC},m}(t) = e^{j(\theta_{\text{OS},m}(t-mT) +\theta_{\text{RF},m} + 2\pi \Delta f_{\text{LOC},m}t - \phi_{\text{OS}}(t) - \phi_{\text{RF}})}
    \label{eq:local_phase}
\end{align}
and
\begin{align}
    \label{eq:ota_phase}
    &\hat{h}_{\text{OTA},m}(t) = v''(t) + \\
    &\rho e^{j(2\pi f_{\text{c},m}(\tau_{\text{p},m} + \tau_{\text{ch}}) + \theta_{\text{OS},m}(t- \tau_m) +\theta_{\text{RF},m} +2\pi \Delta f_{\text{OTA},m}t - \varphi_{\text{OS}}(t) - \varphi_{\text{RF}})},\nonumber
\end{align}
with $v''(t) = v'(t) x_\text{BB}^*(t)$ representing the filtered noise, $\tau_m = mT+\tau_{\text{p},m}+\tau_{\text{ch}}$.
The \ac{CFO}s observed at the local and \ac{OTA} receiver chains are $\Delta f_{i,m} = f_{\text{c},m} - f_\text{c,i}$ with $i \in \{\text{LOC}, \text{OTA}\}$.

\subsection{Phase Estimation}

The phase from \eqref{eq:ota_phase} can be split into the following components (a similar decomposition holds for the local case, but without channel and noise terms):
\begin{itemize}
    \item \textbf{TX Chain phase:} $\theta_{\text{OS},m}(t- \tau_m) +\theta_{\text{RF},m}$, which depends on the specific transmit RF chain $m$.
    \item \textbf{Channel phase:} $2\pi f_{\text{c},m}(\tau_{\text{p},m} + \tau_{\text{ch}})$, resulting from geometry of setup and corresponding propagation delays.
    \item \textbf{CFO at receiver:} $2\pi \Delta f_{\text{OTA},m}t$, separate for each TX chain $m$ as each can have their own synthesizer, resulting in a time dependent phase shift.
    \item \textbf{Noise phase:} $\arg[v''(t)]$, the phase perturbation caused by receiver noise.
    \item \textbf{RX Chain phase:} $\varphi_{\text{OS}}(t) + \varphi_{\text{RF}}$, contributed by the \ac{OTA} receiver oscillator and frontend.
\end{itemize}
It is assumed that RX chain oscillator phases are approximately stationary during transmission of the synchronization signal and that the \ac{CFO}s are negligible.
The phase estimate for chain $m$ is then obtained by the time average over the $m$-th slot of duration $T$ \cite{Kay_Steven10_5555_151045}:
\begin{align}
    \hat{\theta}_{i,m} = \frac{1}{T} \int_{0}^{T} \arg[\hat{h}_{i,m}(\tau)]d\tau.
    \label{eq:phase_est_TX}
\end{align}
Note that under the assumption of sufficiently high \ac{SNR}, the argument can be considered approximately linear, so that the time average yields an unbiased phase estimate.

\begin{figure}[tb]
    	\centering
    	\includegraphics[width=\linewidth]{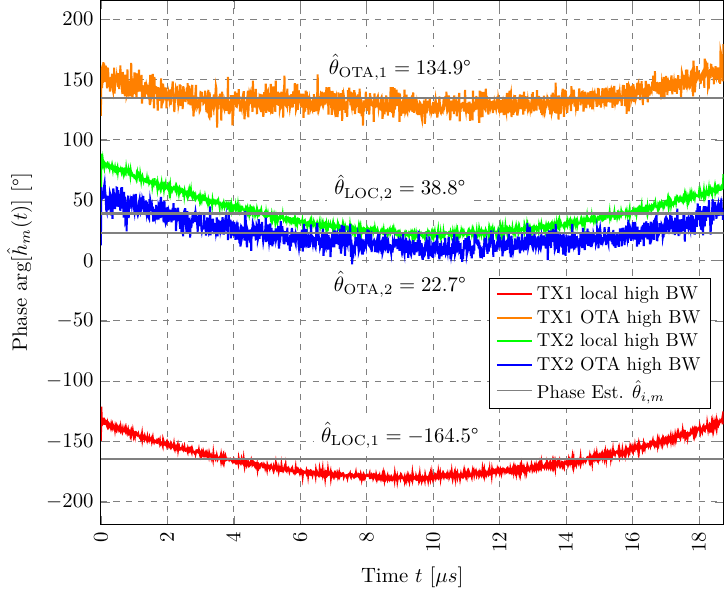}
        \caption{Measured phases over synchronization frame duration for TX1 and TX2 and corresponding phases estimates (gray line) for local and \ac{OTA} receiver with $B=\SI{40}{MHz}$ bandwidth.}
    	\label{fig:4TX_phases}
\end{figure}

\figref{fig:4TX_phases} shows the measured phases for TX1 and TX2 observed at the local \ac{RF} chain $\arg[\hat{h}_{\text{LOC}}]$ (red and green trace) and \ac{OTA} $\arg[\hat{h}_{\text{OTA}}]$ (orange and blue trace) over the interval of the synchronization signal.
The corresponding phase estimates such as $\hat{\theta}_{LOC,1}$ are indicated by the gray traces.
 System parameters are listed in Table~\ref{table:PN_parameters}, and the measurement setup is detailed in the next section.


\section{Measurement Setup}\label{sec:impairment_model}

\begin{table}[tb]
    \centering
    \begin{center}
        \caption{System parameters measurement.}
            \label{table:PN_parameters}
        \resizebox{\linewidth}{!}{
    \begin{tabular}{| l | l | l | l | l | l |} 
        \hline
        Parameter & Symbol & Value & Parameter & Symbol & Value \\
        \hline
        Carrier freq. & $f_\text{c}$ & $\SI{3.75}{GHz}$&Distance & $R$ & $\SI{2}{m}$\\
        Bandwidth & $B$ & $\{2,40 \} \SI{}{MHz}$&Angle & $\phi$ & $\SI{0}{\degree}$\\
        Sample freq. & $f_\text{s}$ & $\{4,80 \} \SI{}{MHz}$&Observations & $L$ & $10000$\\
        TX RF chains & $M$ & $4$&Obs. Interv. & $t_\text{obs}$ & $\SI{50}{ms}$\\
        TX power & $P_\text{TX}$ & $\SI{0}{dBm}$&Samples & $N$ & $1500$\\
        \hline
    \end{tabular}
    }
    \end{center}
\end{table}

 \begin{figure}[tb]
     \centering   
    \includegraphics[width=\linewidth]{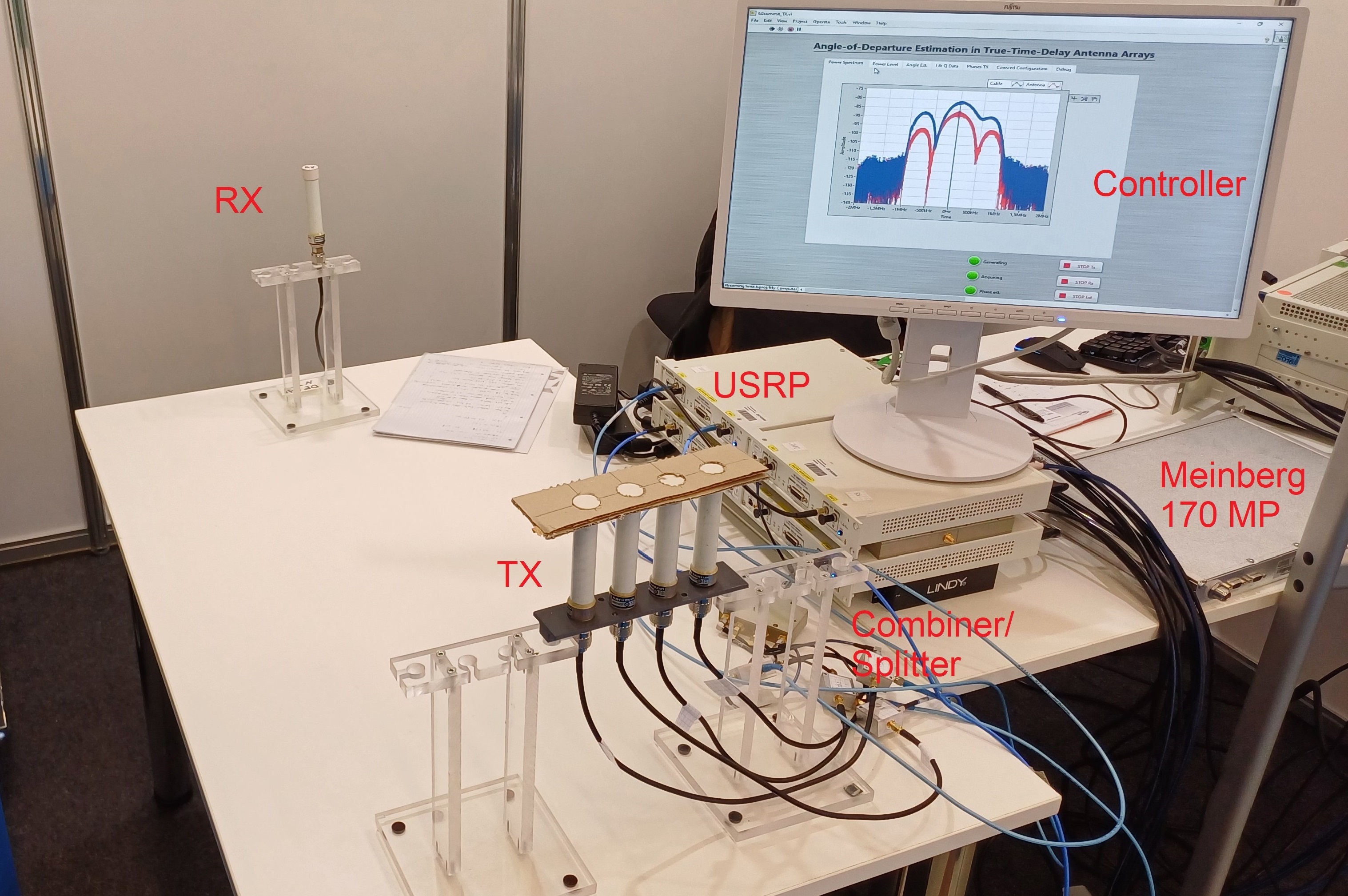}
    \caption{System setup for transceiver with $4$ TX and real-time calibration\cite{Coll202501_2}.}
    \label{fig:schematic_setup}
\end{figure} 
\begin{figure}[tb]
    \centering   
    \includegraphics[width=\linewidth]{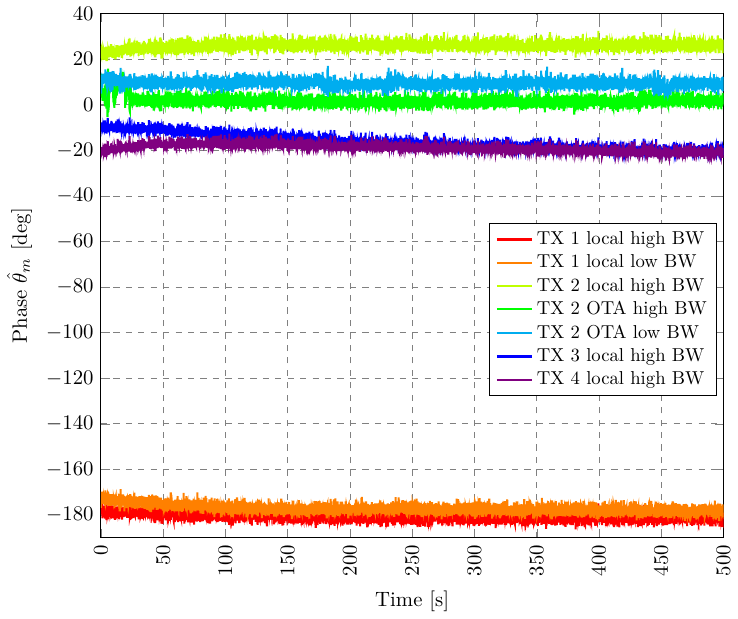}
    \caption{Estimated phases for $4$ transmit chains \cite{1yjg-9863-26} of \ac{USRP}s \texttt{X310} model \texttt{2944R}.}
    \label{fig:TX_phase_meas}
\end{figure}

\subsection{Measurement Hardware Configuration}

The measurement setup, previously described in \cite{Coll202501_2}, is shown in \figref{fig:schematic_setup} and follows the schematic of \figref{fig:TX_sync_setup}.
Its purpose is to measure the phase of each RF chain individually (using a \ac{TDMA} scheme) both \ac{OTA} and through a local wired connection.
Two \ac{USRP} \texttt{X310} model \texttt{2944R}, each providing two \ac{RF} channels, drive the four transmit chains.
Separate \ac{USRP} X310 units of the same model serve as the local reference receiver and the \ac{OTA} receiver, respectively.
All \ac{USRP} are connected to a \ac{GPSDO}, the Meinberg \texttt{170 MP} satellite receiver, which supplies a $\SI{10}{MHz}$ reference and a 1-PPS timing signal for time and frequency synchronization.
Each transmitting chain is connected to a \texttt{ZFRSC-42-S+} splitter by cable.
One output feeds the transmitting array, while the other is combined using a \texttt{ZB4PD1-2000} combiner and routed to the local reference receiver by cable.
The \ac{OTA} signal is captured by a separate receive antenna connected to the \ac{OTA} receiver \ac{USRP}.
Both the TX and RX controllers run on the same \texttt{PXIe-8133} in separate LabVIEW programs. 
Phase estimates computed at the RX controller are sent to the TX controller over a \ac{UDP} socket (feedback).
This allows the TX controller to precode the transmitted signal using either the local reference estimates or the \ac{OTA} phase estimates to achieve coherent transmission.

\subsection{Measurement Procedure and Parameters}

Key system parameters are listed in Table~\ref{table:PN_parameters}.
The $M$ \ac{ULA} elements are spaced at $d$, corresponding to the given carrier frequency.
Measurements are performed at $\SI{2}{MHz}$ and $\SI{40}{MHz}$ bandwidth, to validate the synchronization method from \cite{Coll202606} over a wider bandwidth range.
The transmit power of $\SI{0}{dBm}$ is selected to ensure sufficiently high \ac{SNR} $>\SI{30}{dB}$ for the signals at both the local reference and the \ac{OTA} receiver chain.
The receiver is placed at a known, fixed position at a distance of $\SI{2}{m}$ relative to the transmitting \ac{ULA} at an angle of $\SI{0}{\degree}$ (boresight).
Each transmit chain sends the synchronization signal in its dedicated \ac{TDMA} slot of $N$ samples, repeated periodically every $t_\text{obs}$.
Before and after each chirp, $500$ zero samples are appended as a guard interval.

\subsection{Qualitative Measurement Observations}

\figref{fig:TX_phase_meas} shows the estimated phases from the local reference and the \ac{OTA} receiver for the low and high bandwidth case.
For the sake of brevity, not all the measured phases contained in \cite{1yjg-9863-26} are included.

Two main observations can be made:
\begin{enumerate}
    \item The absolute phase differ slightly between the two bandwidths (red/orange or green/cyan traces), yet the phase remains approximately constant over time with only a minor drift, indicating no significant bandwidth-dependent phase distortion. This confirms that the calibration approach from \cite{Coll202606} extends well to higher bandwidths.
    \item The local and \ac{OTA} phase measurements (lime and green traces) exhibit different phase values, and the \ac{OTA} traces show slightly higher variance. This is consistent with the system models \eqref{eq:local_phase} and \eqref{eq:ota_phase}, which includes channel effects. A quantitative jitter comparison is provided in the following Section~\ref{sec:meas_result}.
\end{enumerate}


While the observed phases remain quasi-stationary over the measurement duration, a slight drift can be seen for the first $\SI{200}{s}$.
For instance the estimated phase of TX3 (blue trace) exhibits a minor drift from $\SI{-10}{\degree}$ to approximately $\SI{-20}{\degree}$, likely caused by warm-up of the \ac{RF} frontend at transmission start.
This is consistent with our previous work \cite{Coll202606}, where it was demonstrated that a minor drift can be effectively compensated by the proposed calibration method.

\begin{figure}[tb]
    \centering
    \includegraphics[width=\linewidth]{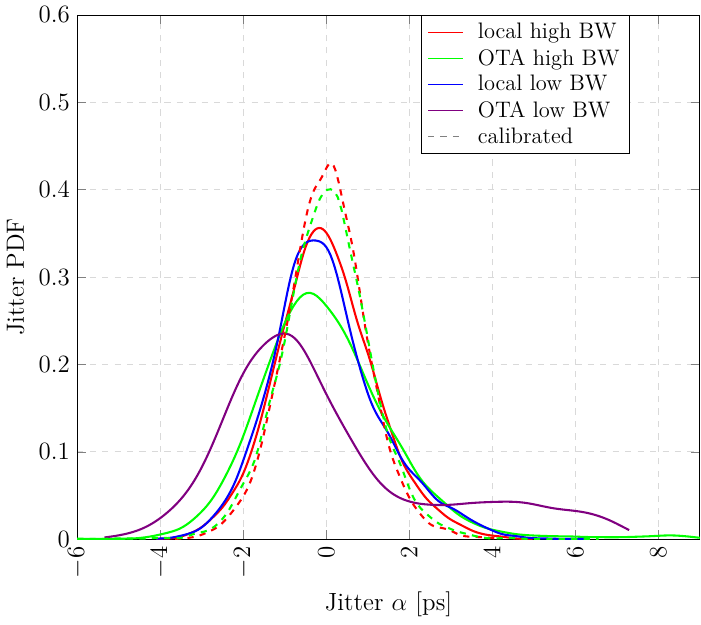}
    \caption{\ac{PDF} for TX1 before/after calibration}
    \label{fig:PDF_KDE_TX1}
\end{figure}
\begin{figure}[tb]
    \centering
    \includegraphics[width=\linewidth]{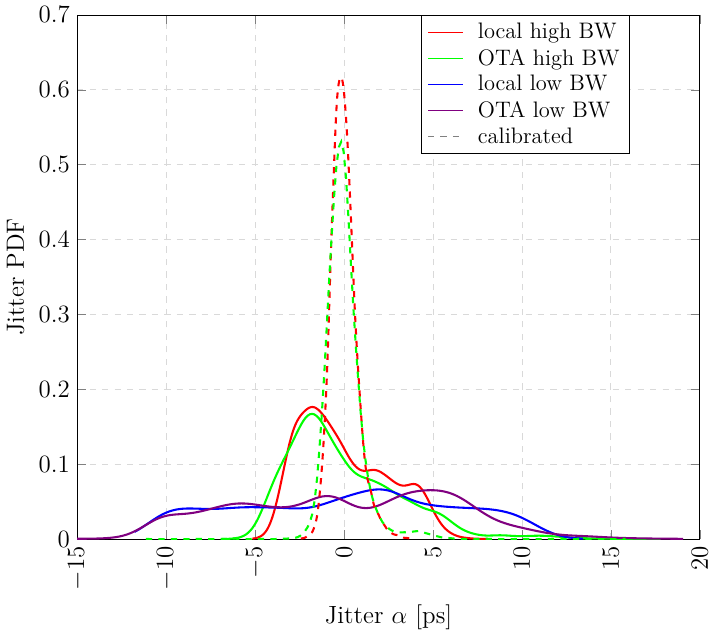}
    \caption{\ac{PDF} for TX3 before/after calibration}
    \label{fig:PDF_KDE_TX3}
\end{figure}

\section{Measurement Results}\label{sec:meas_result}

The phase measurements are conducted periodically in intervals $t_\text{obs}$ for a total of $L$ measurements.
The index $l$ is used to indicate the $l$-th phase estimate of the $m$-th chain $\hat{\theta}_{m,l}$.
Then the corresponding jitter is given by
\begin{align}
\alpha_{m,l} = \frac{\hat{\theta}_{m,l}}{2\pi f_\text{c}}.
\end{align}
The synchronization method considered is the previously introduced \textit{``smoothed calibration''} approach \cite{Coll202606}.
In this method, the transmitted signals are precoded with the average of the last 10 phase estimates, to improve stability.
As a metric to quantitatively assess the synchronization performance, the \ac{RMS} cycle-to-cycle jitter is computed from $\alpha_{m,l}$ (as defined in \cite[(11)]{Coll202606}).

\begin{table}[tb]
    \centering
    \begin{center}
        \caption{\ac{RMS} cycle-to-cycle jitter for different TX.}
            \label{table:jitter_summary}
        \resizebox{\linewidth}{!}{
    \begin{tabular}{| l | l | l | l | l | l | l | l | l |} 
        \hline
        \multirow{3}{*}{} & \multicolumn{4}{c|}{measured} & \multicolumn{4}{c|}{calibrated}\\
                \cline{2-9}
        & \multicolumn{2}{c|}{low BW} & \multicolumn{2}{c|}{high BW}&\multicolumn{2}{c|}{low BW} & \multicolumn{2}{c|}{high BW}\\
        \cline{2-9}
        &Local&OTA&Local&OTA&Local&OTA&Local&OTA\\
        \hline
        TX1 & $\SI{1.31}{ps}$& $\SI{2.52}{ps}$& $\SI{1.17}{ps}$& $\SI{1.61}{ps}$& $\SI{942}{fs}$&$\SI{1.06}{ps}$ &$\SI{980}{fs}$& $\SI{1.08}{ps}$\\
        TX2 & $\SI{1.30}{ps}$& $\SI{2.18}{ps}$& $\SI{1.13}{ps}$&$\SI{1.39}{ps}$& $\SI{951}{fs}$& $\SI{1.01}{ps}$&$\SI{979}{fs}$& $\SI{1.02}{ps}$\\
        TX3 & $\SI{6.19}{ps}$& $\SI{6.24}{ps}$ & $\SI{2.54}{ps}$&$\SI{3.27}{ps}$& $\SI{719}{fs}$&$\SI{1.08}{ps}$ & $\SI{721}{fs}$& $\SI{1.05}{ps}$\\
        TX4 & $\SI{4.71}{ps}$& $\SI{5.45}{ps}$ & $\SI{1.35}{fs}$&$\SI{1.77}{ps}$& $\SI{699}{fs}$& $\SI{839}{fs}$ &$\SI{722}{fs}$& $\SI{772}{fs}$\\
        \hline
    \end{tabular}
    }
    \end{center}
\end{table}

\subsection{RMS cycle-to-cycle Jitter}

Table~\ref{table:jitter_summary} lists the RMS jitter for the four TX chains, obtained from the local reference and the \ac{OTA} receiver at both low and high bandwidths, with and without calibration.
It can be observed that in all cases the calibration yields a significant reduction in the \ac{RMS} jitter.
As a general rule, the measured jitter for \ac{OTA} transmission is higher than at the local reference.
This rule also holds after applying phase calibration.
The explanation for this is the phase from \eqref{eq:ota_phase}, which is impacted by the channel and consequently leading to increased jitter.
In the measurements the jitter values for the low bandwidth case ($\SI{2}{MHz}$) are consistently higher than those for the high bandwidth case ($\SI{40}{MHz}$).
This is because the low bandwidth chirp has a longer duration: the sample rate is lower ($\SI{4}{MHz}$ compared to $\SI{80}{MHz}$), while the number of samples $N$ remains fixed.
A higher measurement duration leads to higher variance of the oscillators phase noise processes $\theta_\text{OS}$, which consequently results in higher jitter \cite{demir_a}.
After applying calibration, the difference between the jitter for the low and high bandwidth cases is negligible.

\subsection{Jitter Probability Density Function}

To qualitatively assess the calibration impact, the \ac{PDF} of the jitter before and after calibration is examined.
For comparison, two \ac{RF} chains are selected: TX1 which exhibits minor drift, and TX3, which exhibits significant drift over the measurement duration.

\figref{fig:PDF_KDE_TX1} shows the \ac{KDE} of the measured jitter values for TX1, before and after calibration.
The measured jitter distributions (red, green and blue trace) are approximately Gaussian with \ac{RMS} values between $\SI{1.17}{ps}$ and $\SI{1.61}{ps}$.
An exception is the measured \ac{OTA} jitter at low bandwidth (violet trace), which exhibits drift and consequently higher \ac{RMS} jitter of $\SI{2.52}{ps}$.
After applying calibration, the \ac{RMS} jitter is reduced, which is graphically visible with lower variance for the dashed traces, compared to the uncalibrated case.
Since the calibrated jitter distributions are nearly identical for the two bandwidths, only one of them is shown in \figref{fig:PDF_KDE_TX1} to avoid redundancy.

\figref{fig:PDF_KDE_TX3} shows the \ac{KDE} of the \ac{PDF} for the measured jitter values of TX3 before and after calibration.
Compared to \figref{fig:PDF_KDE_TX1} it can be seen that the measured jitter is far larger and the \ac{RF} chain exhibits significant drift when no calibration is applied.
In accordance with the measured \ac{RMS} jitter values it can be seen that the low bandwidth (blue and violet trace) case exhibits significantly more drift compared to the high bandwidth (red and green trace) case.
For example a local reference jitter \ac{RMS} of $\SI{6.19}{ps}$ (low bandwidth) vs $\SI{2.54}{ps}$ (high bandwidth) and \ac{OTA} jitter of $\SI{6.24}{ps}$ vs $\SI{3.27}{ps}$ can be observed.
As discussed above, the longer chirp duration in the low bandwidth case allows more phase noise accumulation, leading to larger drift.
After the calibration is applied, the \ac{PDF} becomes approximately Gaussian (dashed traces), and the drift is effectively removed.
Consequently, the \ac{RMS} jitter drops dramatically, for instance in the low bandwidth local reference case from $\SI{6.19}{ps}$ to $\SI{719}{fs}$.

\section{Discussion: Local vs.\ OTA Phase Calibration}\label{sec:evaluation}

The goal of this section is to compare the challenges and practical considerations for local and \ac{OTA} phase calibration methods.
Both calibration methods can use the same core algorithm (such as \ac{TDMA}-based smoothed calibration \cite{Coll202606} or reciprocal calibration \cite{rogalin2015}); the key difference is whether the synchronization signal travels through a cable or \ac{OTA}.
Despite this algorithmic similarity, the two approaches lead to distinct practical trade-offs, as highlighted by our measurements.

\subsection{Local Calibration}

Local calibration offers three decisive benefits due to the absence of a wireless channel:
\begin{itemize}
    \item \textbf{Robustness:} The phase estimates are not affected by multipath, interference, and they do not depend on channel estimation quality.
    \item \textbf{High \ac{SNR} and negligible \ac{CFO}:} The wired connection guarantees high \ac{SNR} and allows the local reference receiver to synthesize its carrier signal from the same $\SI{10}{MHz}$ reference as the TX \ac{RF} chains, eliminating \ac{CFO} in the measurement and minimizing jitter.
    \item \textbf{Low latency and no UE overhead:} The feedback delay is negligible, and all processing resides at the \ac{BS}.
\end{itemize}
The main drawbacks are the need for additional splitters and combiners, and the fact that only the signals up to the antenna connectors are phase aligned.
Potential antenna phase mismatches (observed as negligible in our measurement) must be calibrated separately.

\subsection{Over-the-Air Calibration}

\ac{OTA} calibration avoids separate hardware and directly measures the radiated signals, offering the following advantages:
\begin{itemize}
    \item \textbf{Hardware simplicity:} No extra \ac{RF} distribution hardware is required, making it attractive for large arrays.
    \item \textbf{Antenna alignment:} The complete transmit path (including the antennas) is calibrated, ensuring true beamforming coherence.
    \item \textbf{Integration with channel estimation:} Phase synchronization can be embedded into standard pilot-based channel sounding (e.g., Zadoff-Chu sequences \cite{3gpp_nr_ptrs_chapter_7_4}), reusing existing receiver processing.
    \item \textbf{Location-aware precompensation:} When the \ac{UE} position is known, \ac{OTA} calibration can jointly compensate array phase offsets and channel-induced phase shifts, enabling dynamic beamsteering.
\end{itemize}
A trade-off for these benefits is a higher jitter floor (due to lower \ac{SNR} and channel effects) and a higher feedback latency.

\subsection{Practical Guidance}

Table~\ref{table:jitter_summary} shows that \ac{OTA} calibration consistently yields higher \ac{RMS} jitter than local calibration, especially at low bandwidths where longer chirps accumulate more phase noise.
After applying the calibration, both methods effectively eliminate drift and whiten the residual jitter, but the residual noise level remains lower for the local case.
The choice therefore reduces to a system-level trade-off: local calibration provides superior stability and minimal latency at the cost of extra hardware, whereas \ac{OTA} calibration trades slightly higher jitter for hardware simplicity and antenna phase alignment relevant for beamforming.
For applications where the antenna response is uniform and extra hardware effort is acceptable, local calibration is the more robust option; otherwise, \ac{OTA} calibration offers a practical and easily integrated alternative.

\section{Conclusion}\label{sec:conclusions}

Real-time coherent transmission in \ac{MIMO} systems is impacted by phase impairments such as phase noise and hardware-induced phase offsets.
While receiver-side compensation methods are commonly known, practical transmitter-side phase calibration methods remain underexplored.
In this work, we experimentally compared a local (wire-connected) and an \ac{OTA} phase calibration method.
Both building upon a simple \ac{TDMA}-based smoothed calibration algorithm introduced in previous works, using a fully digital array of \ac{USRP} X310 \ac{SDR}s.

Measurement results across two bandwidths ($\SI{2}{MHz}$ and $\SI{40}{MHz}$) demonstrate that both approaches effectively eliminate phase drift, whiten the residual phase noise, and enable quasi-coherent transmission.
The key insights are:
\begin{itemize}
    \item The local calibration method achieves lower \ac{RMS} jitter and is inherently immune to channel effects, making it the preferred choice when feedback from the receiver is unavailable, latency must be minimized, or the propagation environment is hostile.
    \item \ac{OTA} calibration, while exhibiting slightly higher jitter due to channel impact, eliminates the need for extra hardware and directly calibrates the radiated signals, including the antenna responses. It can be seamlessly integrated into standard channel estimation procedures.
    \item After calibration, the jitter distributions become approximately Gaussian in all tested cases, indicating effective phase-noise whitening and confirming the calibration's robustness for both low and high signal bandwidths.
\end{itemize}

Overall, the experimental evidence confirms that commonly available \ac{SDR} hardware is suitable to implement both local and \ac{OTA} phase calibration, with the choice between them decided by a trade-off between hardware overhead and robustness against channel disturbances.
Future works will extend this approach to scenarios where no joint \ac{GPSDO} for synchronization is available and signals from non-co-located transmitters have to be coherently combined.

    \bibliographystyle{IEEEtran}
	\bibliography{references}

\end{document}